\documentclass{aa} 
\usepackage{graphicx,natbib,amssymb,txfonts}
\bibpunct{(}{)}{;}{a}{}{,} 

\newcommand{\incli}{\theta_{\rm o}}

\begin{document}
\title{The flare model for X-ray variability of NGC 4258}
\author{T. Trze\'{s}niewski\inst{1}
	\and B. Czerny\inst{2}
	\and V. Karas\inst{3}
        \and T. Pech\'{a}\v{c}ek\inst{3}
	\and M. Dov\v{c}iak\inst{3}
        \and R. Goosmann\inst{4}
	\and M. Niko{\l}ajuk\inst{5}}
\institute{Institute of Physics, Jagiellonian University, Reymonta 4, P-30059 Krakow, Poland
	\and Copernicus Astronomical Center, Bartycka 18, P-00716 Warsaw, Poland
	\and Astronomical Institute, Academy of Sciences, Bo\v{c}n\'{\i} II 1401, CZ-14131 Prague, Czech Republic
	\and Observatoire Astronomique de Strasbourg, F-67000 Strasbourg, France
	\and Faculty of Physics, University of Bia\l ystok, Lipowa 41, P-15424 Bia\l ystok, Poland}
\abstract{}{We study the variability mechanism of active galactic nuclei
(AGN) within the framework of the flare model. We examine
the case of Seyfert/LINER galaxy NGC 4258, which is observed at high
inclination angle and exhibits rapid fluctuations in its X-ray
light curve.}{We construct a model light curve based on the assumption
of magnetic flares localized in the equatorial plane and orbiting with
Keplerian speed at each given radius. We calculate the level of
variability as a function of the inclination of an observer, taking into
account all effects of general relativity near a rotating supermassive
black hole.}{The variability level is a monotonic function of the source 
inclination. It rises more rapidly for larger values of the black
hole spin (Kerr parameter $a$) and for steeper emissivity (index $\beta$
of the radial profile). We compare the expected level of variability for 
the viewing angle $81.6$~deg, as inferred for NGC 4258, with the
case of moderate viewing angles of about $30$~deg, which are 
typical of Seyfert type-1
galaxies.}{Highly inclined sources such as this one are
particularly suitable to test the flare model because the
orbital motion, Doppler boosting, and light bending are all expected to 
have maximum effect when the accretion disk is seen almost edge-on. The model
is consistent with the NGC 4258 variability, where the obscuring material
is thought to be localized mainly toward the equatorial plane rather 
than forming a geometrically thick torus. Once the intrinsic
timescales of the flare duration are determined with higher precision,
this kind of highly inclined objects with a precisely known mass of 
the black hole can be used to set independent constraints on the spin 
parameter.}
\keywords{accretion, accretion disks -- galaxies: active -- galaxies: Seyfert -- X-rays: galaxies}
\date{Received 20 December 2010; Accepted 18 April 2011}
\authorrunning{Trze\'{s}niewski et al.}
\titlerunning{The flare model for X-ray variability of NGC 4258}
\maketitle

\section{Introduction}
\label{sec:introduction}
For many years the X-ray emission of active galactic nuclei (AGN) has
been known to be strongly variable \citep[see e.g.][for
reviews]{gaskell06,uttley07}. However, the nature of this variability
remains unknown. Most proposed models have been based on a natural
assumption that the emission originates close to the accreting
supermassive black hole and fluctuates on the
dynamical timescale. Rapid variability is among the arguments in
favor of this interpretation \citep{krolik99}. The effects of general
relativity are expected to play an important role in shaping the
observed signal \citep{kato98}.

As the innermost part of accretion flow likely proceeds through some
form of a disk, characterized by a (roughly) Keplerian profile of
rotational velocity, the relativistic effects are expected to depend on the
inclination angle of the observer with respect to the disk plane. 
These effects are only weakly seen in the stationary continuum 
models if the emission is due to Comptonization, but they are very
profound in the observed shapes of spectral features such as 
the iron K$\alpha$ line (e.g. \citeauthor{fabian95} 
\citeyear{fabian95}).  In addition, earlier
studies predict that non-stationary continuum models can display 
relativistic effects also through the dependence of the level 
of variability on the inclination angle of an observer
 \citep{zhang91,abramowicz91,karas98,fukue03,czerny04}.
Therefore, the application of a specific model to both low and high
inclination sources opens an additional possibility of model testing.

In the present paper, we apply the flare model developed originally for
the case of Seyfert galaxy MCG--6-30-15, which is a source at a moderate
inclination ($\incli\sim30$~deg) with respect to the observer's line of
sight (\citeauthor{fabian95} \citeyear{fabian95}, \citeyear{fabian02}),
to the \object{NGC 4258} galaxy viewed from the almost edge-on direction. This
active galaxy is unique in several aspects: well-resolved maser
emission from the nucleus allows an accurate mass determination of the
black hole; the source is Compton thin (despite the high inclination),
which allows us to measure the X-ray variability; rotation studies also
show that the inner accretion disk follows Keplerian orbital motion
very accurately.

The paper is organized as follows. In Sect.~\ref{sec:application}, we
summarize the general scenario of the flare model, introduce a 
convenient parameterization, and adapt this scheme to the case
of NGC 4258. We discuss the main differences in the model set-up
caused by our concentration in previous papers
on low to moderate inclinations for the application to 
unobscured Seyfert~1 AGN, whereas now we wish to apply the model
to a highly-inclined source. In Sect.~\ref{sec:results}, we
present our showing in particular the inclination dependence
of the variability variance. We consider the role of the avalanche 
prescription, where the flares are mutually interconnected as they 
occur in families. Finally we summarize the results in 
Sect.~\ref{sec:discussion} and present our conclusions.

\section{Flare model for X-ray variability of NGC 4258}
\label{sec:application}

\subsection{Description of the flare model}
\label{sec:model}
The idea of X-ray emission coming from multiple locations within/above
accretion disks around a black hole is motivated by the important role
of magnetic fields in the process of accretion. The original 
formulation of the flare model by \citet{galeev79} was followed 
by numerous papers
\citep[e.g.][]{dimatteo98,nayakshin01,merloni01,collin03,goosmann07b}.
A common theme to these models is an underlying assumption of a standard-type
accretion flow driving the magnetic field. A similar picture can also be
developed in the absence of a cold disk since in that case multiple
shocks are expected to form in the hot accretion flow
\citep[e.g.][]{boettcher99,zycki03}. The attractiveness of the flare
model is also supported by the close correlation between the
radio and X-ray emission in radio quiet AGN, which is a phenomenon
typical of active stellar coronae \citep{laor08}.

In the present work, we generalize the model developed by
\citet{czerny04} and \citet{goosmann06}. We parameterize the distribution
of the flares across the disk and the flare properties. We assume that
the luminosity of a single flare following the initial onset decreases
gradually with time. This differs from the previous paper where
we adopted a rectangular profile for the flare light curve. The most
important modification is the introduction of a coupling between the
flares through the presence of avalanches, as discussed e.g. by
\citet{poutanen99}, \citet{zycki02}, and by \citet{pechacek08}.
Thus, our model follows the general idea of the scheme based on the
propagation of perturbations
\citep{mineshige94,lyubarskii97,kotov01,arevalo06,titarchuk08}.

Spontaneous flares originate at a certain outer radius,
$R_{\rm{}spont}$. Its value can be rather distant from the center, as
this seems to follow from the discussion of well-studied cases of
major flares in AGN \citep[e.g.][]{ponti04,goosmann07a}. 
On a more theoretical side, \citet{uzdensky08} described 
a specific scenario of a magnetized 
corona above a turbulent accretion disk. Magnetic loops rise above
the disk, where they become sheared by differential rotation of the
field line foot-points until reconnection occurs. These events are 
responsible for the occurrence of primary flares and the
subsequent dissipation of magnetic energy.
 
Spontaneous flares are accompanied by avalanches of secondary flares
developing subsequently inside $R_{\rm{}spont}$. The assumptions about 
the parent flare localization and the gradual
avalanche progression towards smaller radii go beyond the standard
avalanche model, which does not invoke any particular form of the
spatial distribution. General arguments,
based mainly on the diffusion equation for the propagation of the
events, suggest that the properties of secondary flares are described
by power-law distributions. In this way, the onset of variability
occurs at large radii and induces secondary fluctuations further
down the accretion flow, in the inner regions of the disk.

In our paper, the flares are assumed 
to vanish at the inner radius of the computational zone, 
$R_{\rm{}in}$, which corresponds to the final inflow of 
perturbations below the innermost stable circular orbit (ISCO), 
and eventually below the black hole horizon. The assumption about 
the exponential decay of individual flares is another improvement of our
previous scheme \citep[see][where we assumed a common
characteristic lifetime for the duration of all flares which then
switch off abruptly]{czerny04}. Onset of flares followed by
their rapid but gradual decay may be analogous to
processes in solar flares, where similar mechanisms of heating via magnetic 
reconnection operate \citep[e.g.,][]{aschwanden98}.

\subsection{Parameterization of the model}
We briefly summarize the adopted parameterization. Flare events are
characterized by radius from the black hole and the corresponding time
of the occurrence. During their lifetime, the flares follow a coplanar
Keplerian motion of the accretion disk. 

As we wish to take general relativity effects into account, the flare
event coordinates need to be defined in an appropriate manner. For
the purposes of our investigation, the Boyer-Lindquist $(t,r)$ coordinates
suit well, because all flares are assumed to be distributed above the
ISCO, so these coordinates are non-singular over the whole computational
domain. An individual flare is located at the radius $r$ and
characterized by the moment of its birth, $t_{\rm{}birth}$. The flares
rise instantaneously and decay with time $t$ exponentially (with a sharp
final cut-off)
\begin{equation}
f(t) = \left\{\begin{array}{ll}
f_{1}(r)\,\exp\left(-\epsilon\;\frac{\textstyle{t - t_{\rm{}birth}}}{\textstyle{\tau(r)}}\right), & t_{\rm{}birth} < t < t_{\rm{}birth} + \tau(r), \nonumber \\~&\\ \hfill
0,\hfill & t > t_{\rm{}birth} + \tau(r)\;\mbox{or}\; t<0.
\end{array}\right.
\label{eq:tbirth}
\end{equation}
The amplitude $f_1(r)$ of a flare depends on the flare location, and
its dependence on the radial coordinate has the form
\begin{equation}
f_1(r)=f_{0}\;\left({r \over r_0}\right)^{-\beta},
\label{eq:beta}
\end{equation}
where $f_{0}$ is a normalization constant and $\beta$ is a model
parameter. The lifetime of the flare is assumed to be related to the
flare location as
\begin{equation}
\tau(r)=\tau_{0}\;\left({r \over r_0}\right)^{\delta},
\label{eq:tau}
\end{equation}
where $\tau_{0}$ and $\delta$ are the model parameters. The scaling radius
$r_0$ was set  at $18R_{\rm{}g}$. 

In our picture, we assume that each flare can lead to a new flare with
the probability given by a Poissonian distribution around a mean value
$\mu$. Each new flare is always generated closer to the black hole than
the parent flare: the location of the new flare, $r_{\rm{}new}$, is
derived from the radii between the parent flare radius, $r$, and the
radius $sr$ ($0\leq s < 1$), where the radial distribution of the probability
is defined as 
\begin{equation}
p(r_{\rm{}new}|r)=n_{0}(r)r_{\rm{}new}^{\gamma+1},
\label{eq:gamma}
\end{equation}
where $n_{0}$(r) is the local normalization constant depending on the
position of the parent flare, and $\gamma$ is a model parameter. The value
$\gamma = 0$ describes the uniform distribution across the disk surface.
The time of birth of a new flare is drawn uniformly from the time bin
between the birth of the parent flare, $t$, and $f_{\rm{}delay}t$, where
the dimensionless factor $f_{\rm{}delay}$ is of the order of a few units.

The level of variability is set by global parameters: the number of
spontaneous flares, $N_{\rm{}f}$, and the total duration of the
light curve, $T_{\rm{}tot}$. The size of the single emitting region is
then identical for all flares. It is fixed by the number of flares
present in the source and the average luminosity $L$ of the source. 
Flares are in Keplerian motion around the black hole. All general
relativity effects connected with light propagation toward an observer
are calculated using the {\sc ky} code \citep[for details
see][]{dovciak04a,dovciak04b}. For that purpose, all emitting regions are
assumed to be projected onto the equatorial plane.

The shape of the locally emitted spectrum is parameterized as a power
law of a given slope, and the reflection component with arbitrary 
normalization is usually included in the model \citep{czerny04,goosmann06}.

\subsection{Properties of NGC 4258}
\label{subsec:properties}
NGC 4258 (M 106) is a barred spiral galaxy with a low-luminosity, type 1.9 
Seyfert nucleus, which is also classified as a LINER.\footnote{See The NASA/IPAC 
Extragalactic Database (NED), which is operated by the Jet Propulsion Laboratory, 
California Institute of Technology, under contract with the National Aeronautics and 
Space Administration.} The nuclear source has a luminosity of $L\approx
10^{-4}L_{\mathrm{Edd}}$ \citep{fruscione05} and
$L_{2-10}=3.31\times10^{40}$~erg/s in the $2$--$10$ keV band. 
The geometrical distance to NGC~4258 can be inferred from
the orbital motion of H$_{2}$O masers in its nucleus, as demonstrated 
for the first time by \citet{herrnstein99}, obtaining
$(7.2\pm0.3)$~Mpc. The mass measurement is reliable because the maser
rotation curve is most accurately described by the Keplerian profile \citep{hern95}. 
Since the radius of the inner masers, measured with the VLBI, corresponds to
$2.8$~mas, high-velocity masers yield a mass for the central black hole
of $M=3.78\times10^{7}M_{\odot}$ \citep{martin08}. The corresponding
Eddington luminosity is $L_{\mathrm{Edd}}=47.5\times10^{44}\
\frac{\mathrm{erg}}{\mathrm{s}}$, hence $L\approx 0.00475\times 10^{44}$~erg/s. 

The masers indicate that the angle of inclination of the inner disk is
high with respect to the observer line of sight:
$\incli\approx81.6^{\circ}$, where $90^{\circ}$ represents edge-on
orientation \citep{martin08}. \citet{wilkes95} find that optical lines
are strongly linearly polarized and that the position angle coincides with
the plane of the maser disk. Intrinsically, the source is likely to be a
Seyfert 2 galaxy because the emission lines seen in the polarized light
have widths appropriate for this class of objects. 

The source was classified in X-rays as an obscured low-luminosity active 
nucleus by the ASCA observation \citep{makishima94}. It 
was observed at X-ray wavelengths by two high-spatial resolution
instruments: Chandra \citep{young04} and XMM-Newton \citep{pietsch02}.
More recently, results derived from Suzaku observations have also been reported
\citep{reynolds09,yamada09}. The nuclear emission
is strongly variable \citep{markowitz05,reynolds09,yamada09}. It follows
that the surrounding medium cannot be Compton-thick along the line of
sight from the entire X-ray producing volume. The obscuration instead
causes a partial absorption with the absorbing column $\sim 10^{23}$
cm$^{-2}$ \citep{yang07,yamada09}. The obscuring material can be
localized mainly towards the equatorial plane rather than form a
geometrically thick torus. 

The slope of the X-ray spectrum
is typical of Seyfert galaxies, where the value $\Gamma \sim 1.7$--$2.0$
depends on the model details \citep{pietsch02,yang07}. The cold and
narrow iron line comes from the outer region, whereas no broad line was
required to fit the XMM-Newton spectrum of this source
\citep{yang07,reynolds09}. \citet{yamada09} confirm a significant X-ray
variability and conclude that a classical geometrically thick torus
does not seem to be present in this object, but that instead, the obscuration may
take place along the line of sight close to the equatorial plane 
\citep[e.g.,][]{bao92,karas92}.

\subsection{X-ray excess variance in NGC 4258 and the enhancement ratio}
The X-ray flux of NGC 4258 is highly variable. The level of this variability
is conveniently characterized by the dimensionless excess 
variance \citep{Nandra97}
\begin{equation}
({\tt{}rms}/{\tt{}mean})^2 = {1 \over N\mu^2} \sum^N_{i=1}[(X_i - \mu)^2 - \sigma_i^2],
\end{equation}
where $N$ is the number of (equally spaced in time) data points, $X_i$ is the 
count rate in a given time bin, $\mu$ is an unweighted arithmetic mean of 
$X_i$, and $\sigma_i$ is the measurement error (predominantly consisting of 
Poissonian noise). The error level depends on the duration of the light curve.
This dependence
is a typical property of the sources dominated by the red noise, as discussed 
for example by \citet{vaughan03}.

The source was observed several times with the ASCA satellite. These data were
used by \citet{nikolajuk09}, who obtained
\begin{equation}
({\tt{}rms}/{\tt{}mean})^2_{\rm observed}=6.215^{+1.566}_{-2.350}\times10^{-3}
\label{eq:var_value}
\end{equation}
for several light curves of typical total duration  $T=31\,300$~seconds,
consisting of  $50$ bins and the duration time of a single bin 
$t_{\rm{}b}=626$~s.

Since the black hole mass in NGC 4258 is known, this value can be compared
to the expected variance from the relation between the black hole mass and
the X-ray excess variance, that is characteristic of Seyfert galaxies 
\citep{nikolajuk06,nikolajuk09}
\label{eq:mass_formula}
\begin{equation}
({\tt{}rms}/{\tt{}mean})^2 =  C\; {T - 2\,t_{\rm{}b} \over M},
\end{equation}
where $C = 1.92 M_{\odot}$ s$^{-1}$.

For the black hole mass $3. 78 \times 10^7M_{\odot}$ 
(see Sect.~\ref{subsec:properties}) and the light curve parameters $T$ and 
$t_{\rm{}b}$ as above, we obtain
\begin{equation}
({\tt{}rms}/{\tt{}mean})^2_{\rm expected} = 1.52 \times 10^{-3}. 
\end{equation}

We thus define the enhancement ratio
\begin{equation}
{\cal R} =  { ({\tt{}rms}/{\tt{}mean})^2_{\rm observed} \over ({\tt{}rms}/{\tt{}mean})^2_{\rm expected}},
\end{equation}
and in the case of our source this ratio is equal to
\begin{equation}
\label{eq:ratio_4258}
{\cal R} = 4 \pm 1.
\end{equation}
This (rather large) error arises mainly from the systematic error in the 
constant $C$ in Eq.~\ref{eq:mass_formula}.

We expect that the enhancement is due to the inclination of the source,
which in the case of NGC 4258 is significantly larger than the 
average inclination (about $30$ deg) of sources usually
used to derive the Eq.~(\ref{eq:mass_formula}) \citep[see e.g.][]{nikolajuk06}. 

\subsection{Adaptation of the model to NGC 4258}
\label{subsec:adaptation}
In our calculations, we used values of the black hole mass and an AGN
luminosity appropriate for NGC 4258 given in the previous subsection.
For the physical structure of the model, because of the lack of hard
evidence of a disk in the vicinity of a black hole in
the NGC 4258 nucleus, we assumed that only primary
emission, from the flares, is present (no cold accretion disk,
therefore, no spots are induced by flares on the disk surface). We assume that
this emission is concentrated all around the equatorial plane.

Some model parameters were constrained to obtain a
profile of the X-ray power-spectral density that is close to that for
MCG--6-15-30 as a general representation of the observed shape for AGN.
The values of the model parameters that we did not vary are as follows:
$\delta=1.5$ (assuming that flare duration scales with local Keplerian timescale),
$\gamma=1.5$, $ s = 0.2$, $f_{\rm{}delay} = 7$, $\mu = 1.1$, and
$R_{\rm{}out} = 200$. Dimensionless distances are given in units of the
black hole gravitational radius, $R_{\rm{}g}\equiv\,GM/c^{2}$. Our model 
of course allows us to change the above-mentioned parameters if the physical
reasoning requires us to use different  values. We tested the model
dependence on the luminosity concentration  given by $\beta$ (see
Eq.~\ref{eq:beta}), on the proportionality constant of the flare
duration, $\tau_0$ (see Eq.~\ref{eq:tau}), and on the Kerr parameter, $a$.

The quantities in physical units, namely, the mass, spin, and distances, 
are related to the corresponding values in geometrical units
\begin{equation}
\frac{M^{\rm phys}}{M^{\rm phys}_{\odot}}=
     \frac{M}{1.477\times 10^{5}{\rm cm}}, \quad 
  a^{\rm phys}=ca,\quad r^{\rm phys}=r.
 \end{equation}
Furthermore,
\begin{equation}
\frac{a}{M}=\frac{a^{\rm phys}}{GM^{\rm phys}/c} \,,\quad
  \frac{r}{M}=\frac{r^{\rm phys}}{GM^{\rm phys}/c^{2}} \,.
\end{equation}
We employ geometrical units in numerical simulations, but convert them
to physical units interpreting the results. To obtain
frequency in physical units [Hz], the relation is $\omega^{\rm
phys}=c\omega$. The geometrized frequencies scale as $M^{-1}$. Their
numerical values must thus be multiplied by a factor 
\begin{equation}
\frac{c}{2\pi
M}=(3.231\times 10^{4})\, \left(\frac{M}{M_{\odot}}\right)^{\!-1}\,[{\rm{Hz}}].
\end{equation}
To calculate the light curves, we followed the ASCA setup, as
the source was observed several times with this satellite. This helps
us to determine the excess variance to a
sufficiently high precision \citep{nikolajuk09}.
Therefore, in our calculations we applied the total
duration of the observation $T=31\,300$~seconds with $50$ bins and
the duration time of a single bin $t_{\rm{}b}=626$~s.

Since the X-ray spectrum of NGC 4258 does not show the reflection 
component, the reflection must be either weak or absent. Therefore, in the present 
study we simplify the model and neglect the reflection. In this case, the 
additional parameter related to the normalization of the reflected component 
is unnecessary. The results in this case even more remarkably 
do not depend on the assumed slope of the power law emission.

For all model light curves, we calculated the variance and then tested its
dependence on the inclination angle, while varying the Kerr parameter of the
black hole and the radial distribution of the flare luminosity, $\beta$.
The latter determines whether the emissivity is more or less concentrated
toward the black hole. 

\section{Results: Dependence of variance on inclination}
\label{sec:results}
We use the aforementioned model to calculate the
normalized X-ray excess variance, $({\tt rms}/{\tt mean})^2$, paying
special attention to the relation between this quantity and the observer
inclination. The flare model variability is very sensitive to the
inclination, so we first explore this dependence while several other
model parameters are kept fixed. In particular, we set the black hole
mass to the appropriate value for NGC 4258. As for the black hole
rotation, since we have no constraints for the spin parameter in NGC
4258, we also keep it free and analyze the level of variability for
different values of $a$. 

\begin{figure}
\centering
\scalebox{0.7}[0.7]{\rotatebox{-0}{\includegraphics{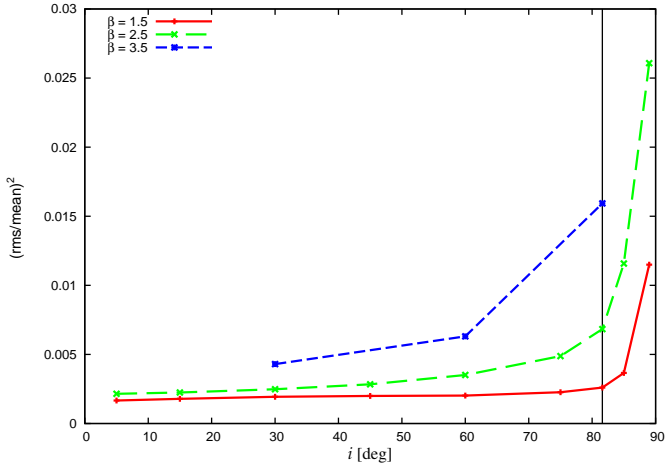}}} 
\caption{The dependence of the normalized variance on the inclination
angle of an observer for three emissivity profiles $\beta$ for
the Kerr black hole $a=0.95$. The vertical line indicates the inclination 
of $\incli=81.6^{\circ}$.}
\label{fig:a95}
\end{figure}

\begin{figure}
\centering
 \scalebox{0.7}[0.7]{\rotatebox{-0}{\includegraphics{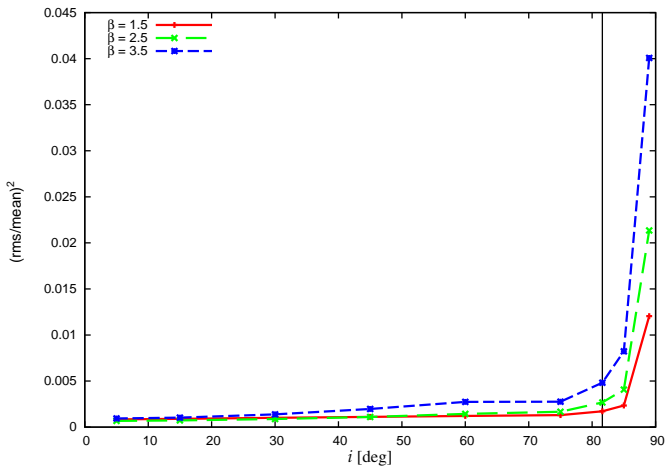}}} 
\caption{The dependence of the normalized variance on the inclination
angle for the same values of $\beta$ as in the previous figure, but 
with a non-rotating (Schwarzschild, $a=0$) black hole.}
\label{fig:a0}
\end{figure}

In Figure~\ref{fig:a95}, we plot the X-ray variance for three values of
$\beta$ assuming a rapidly rotating black hole (Kerr parameter $a = 0.95$).
The value of  $\beta\simeq1.5$ corresponds to a shallow emissivity
profile. The variability level increases with the inclination, first
slowly and later sharply. The rise is the result of the beamed Doppler
boosting at high inclinations, which is well-known from previous papers
\citep[e.g.][]{czerny04,dovciak08}. The rise is faster for larger values 
of $\beta$, which implies that there is a larger contribution to the total 
luminosity of the source arising in the inner region. As a 
consequence, the relativistic effects are strongly enhanced.

The aforementioned increase in the variance with inclination is slower and
the overall  variability level is lower if the black hole is not
rotating. In Figure~\ref{fig:a0}, we show the inclination  dependence of
the normalized variance for the same set of radial profiles $\beta$ as
in Fig.~\ref{fig:a95}, but now with the Schwarzschild black hole. The
ratio of the variances for $\beta = 3.5$ at $\incli = 81.6^{\circ}$ and 
$\incli=30^{\circ}$ is equal to $3.5$ for an $a = 0$ black hole, but increases to
$3.7$ for an $a = 0.95$ Kerr black hole in the presented plots. 

The rise in the variance with the inclination angle is not a result of the
statistical errors in simulations. It is well known that a single
simulation of short timescale is strongly affected by the
power leaking from the lower frequencies
\citep{vaughan03}. However, in Figs.~\ref{fig:a95} and \ref{fig:a0} 
the sequences for  fixed values of $\beta$ were calculated for the same
realizations of the same  statistical distribution. The rise in variance is entirely
due to the change in the viewing angle. If we use different random
realizations of the flare distribution the  trend is less clear because of
large statistical dispersion in the adopted length of the light curve. A
single variance (in simulations as well as in the actual data) is
determined with an accuracy of a factor of two. 

\begin{figure}
\centering
 \scalebox{0.7}[0.7]{\rotatebox{-0}{\includegraphics{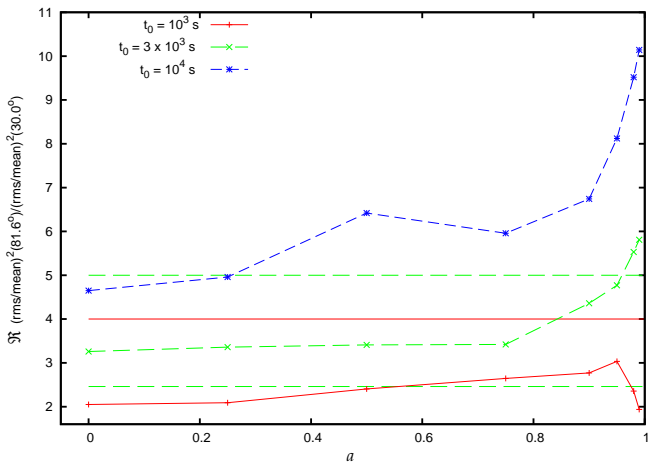}}} 
\caption{The ratio $\cal R$ of normalized variance at the observer inclination 
angle $\incli = 81.6^{\circ}$ with respect to the case  $\incli = 30^{\circ}$.
The dependence on the dimensionless Kerr black hole 
parameter $a$ is plotted for the assumed 
emissivity profile $\beta = 3.5$, for three values of the flare lifetime scale $t_0$, as 
indicated in the plot.  Three horizontal lines give the enhancement ratio $\cal R$ 
for NGC 4258 given in Eq.~(\ref{eq:ratio_4258}) (middle line), 
and the range of errors (upper and lower lines). The simulated light curves were 
longer by factor 8 compared to those used in Figs.~\ref{fig:a95}--\ref{fig:a0}.}
\label{fig:ratio}
\end{figure}

Therefore, to show the model predictions to higher accuracy,
we had to extend the simulated light curves by a factor of eight.  For these
longer light curves, we calculated the variance enhancement, defined as
the ratio of the variance seen at $81.6^ \circ$ to that at $30^
\circ$, as a function of the Kerr parameter. The result is shown in
Fig.~\ref{fig:ratio}, for three values of the flare time-scale duration.
We also plot the ratio of the observed variance, given in
Eq.~(\ref{eq:var_value}), to the variance expected from Eq.~(4) of
\citet{nikolajuk09}. This ratio is found to equal $4.0$, its standard deviation
($1\sigma$) error coming from the errors in both the variance and in
scaling constant. Because of a lack of information about the black hole
spin, the observational constraint is indicated by straight lines
irrespective of the value of $a$. 

The variance enhancement, in general, increases with the Kerr parameter but
the detailed profile depends on the assumed scaling factor of the flare
lifetime. The variability is yet greater when the flare lifetime
close to  the black hole is comparable to the time spent by the flare in
the region of the highest Doppler boosting. The angular extension of
this region (as a fraction of $2 \pi$) decreases as the
Kerr parameter increases \citep{dovciak04a}.

The enhanced variability is consistent with the expectations of the
relativistic enhancement, within the framework of the flare model. A 
flare duration of longer
timescale, $\tau_0 = 10^4$~s, is indicative of a
non-rotating black hole, while the flares scaled down to $\tau_0 =
10^3$~s probably correspond to a spinning black hole with $a > 0.5$. However, taking
into account the large error in the observed variance and the lack of a
priori knowledge of the flare timescale we cannot at this stage make any
firm conclusion about the black hole rotation in NGC~4258.

\section{Discussion}
\label{sec:discussion}

In the flare model of AGN X-ray variability the change in the X-ray flux
is caused by a combination of two main effects. The first is the intrinsic
variability of both a single flare and the flare distribution. The second
is the variation caused by the relativistic effects of the
flare orbital (Keplerian) motion. Analyzing the change in the level of
variability with the inclination we can disentangle those two
effects.

We chose NGC 4258 as a target of our study because of its exceptional
properties. The source is highly inclined, seen almost edge-on, but is
nevertheless Compton thin and the variable primary emission is
still transmitted effectively through the material located at the source
equatorial plane, although the X-ray emission is significantly absorbed.
In addition, the mass of the central black hole in this object is
measured accurately thanks to the spatially resolved water maser emission.

On the other hand, we modified our original flare model \citep{czerny04} by including the
exponential decay of individual flares and the effect of an
avalanche-type development of flares. On the other hand, we neglected
the cold disk reflection. In NGC 4258, any manifestations of reflection
features, such as the broad relativistic iron line and temporary
spots on the disk surface, are strongly suppressed. We thus understand
the situation by presuming that the central corona is the source of
enhanced variability and the place where flares originate.

After developing a refined version of the model, we analyzed the dependence of
the X-ray variance on the inclination angle. We noticed that the rise 
in the variability
with the inclination angle is faster if the emissivity is more
concentrated towards the black hole, which is the case for larger values
of the $\beta$ parameter. We compared the expected level of variability
for the viewing angle $81.6$~deg, as inferred for NGC 4258, with the
case of moderate viewing angles $\incli\approx30$~deg, which are typical of Seyfert~1
galaxies \citep{Nandra97b,Nandra2007}. Our model predicts that the
variability increase is monotonic with inclination.

The observed variability level in NGC 4258 is indeed higher by a factor
consistent with the measured value of the central black hole mass, which
was determined with high accuracy in this source from water maser
emission. The measured X-ray excess in this source is equal to
$\sigma^{2}=6.215^{+1.566}_{-2.350}\times10^{-3}$ \citep{nikolajuk09}.
The expected variance from the Eq.\ (4) of \citet{nikolajuk09} is a
factor of four lower (the aforementioned formula is satisfactory for typical
low-inclination Seyfert galaxies). This means that the flare model of
variability would be consistent with this object provided that the X-ray
variance for a highly inclined source is a factor of four higher than it is
for low-inclination sources for which the relation was scaled.

Several parameters determine the outcoming light curve in our model.
The interplay between the parameters is in general quite complex
because the signal depends on the initial distribution of parent
flares, their subsequent propagation across the accretion
disk towards the inner edge, as well as the avalanche mechanism
generating the secondary flares. To capture this entire 
process, we introduced the flare rise time $t_{\rm birth}$ in 
eq.~(\ref{eq:tbirth}), the index of radial distribution of flare 
amplitudes $\beta$ in eq.~(\ref{eq:beta}), the radial profile of 
flare lifetime $\delta$ in eq.~(\ref{eq:tau}), and the index of 
radial probability distribution $\gamma$ in eq.~(\ref{eq:gamma}). 
Furthermore, the secondary flare generation is described by the
parameters $s$, $f_{\rm{}delay}$, $\mu$, $R_{\rm{}in}$, and
$R_{\rm{}out}$, which define the development and the gradual decay
of the avalanches.

Despite the complicated picture given by the number of free 
parameters described above, it is interesting
to note that only in special (but quite representative) cases do some 
parameters play an important role. This is in particular the case of 
a stationary distribution that is expected to arise when the secondary
flares occur over the whole range of radii in the accretion disk 
($s\rightarrow0$). One can then check that the stationary situation depends
sensitively on $\mu$ (i.e., the mean value of secondary flares) 
but the degree of the dependence on other parameters is much weaker.

\section{Conclusions}
The above-mentioned rise of variability agrees with the model
prediction for a steep  emissivity profile and the flare normalization
timescale of $10^3$--$10^4$~s.  Higher values of the flare timescales
are consistent with the low values of the black hole spin in NGC 4258,
whereas shorter
timescales are consistent with a rapidly spinning black hole.

The observed enhancement in variability supports the view that the X-ray
emission is generated close to the black hole and subject to strong
general relativity effects. This does not imply that the emission
originates preferentially from magnetic flares above the disk, or that
it instead comes from individual shocks in an optically thin inner flow
that undergoes Keplerian motion. In the absence of a reflection component,
our model is unable to differentiate between the two aforementioned scenarios.
However, if the inflow is almost spherical, as in the models of
\citet{shrader98} and \citet{ishibashi09}, the variability enhancement
is not expected for high inclination sources. The observed enhancement of 
variability of NGC 4258 thus implies that there is a significant azimuthal motion in
the X-ray emitting plasma of the inner accretion disk. 

Once the intrinsic timescales of the flare duration are determined with
higher precision, the highly inclined objects with precisely known black hole 
masses can be used to set independent constraints on the
spin parameter. The highly inclined AGNs that are not Compton thick and at
the same time exhibit maser sources, such as NGC 4258, do exist (even
though they are rather rare) and should be ideal candidates to 
apply the method described in this paper. We will however need to 
obtain a better coverage of their variability properties. It appears that
unabsorbed Seyfert 2 galaxies \citep{brightman08} or even dust-poor
quasars \citep{hao10} form a category of sources that could be
suitable candidates.

\begin{acknowledgements}
This work was supported by grants 1P03D00829 and NN~203~380136 of the
Polish Ministry of Science and Education, and the grant ME09036
of the Czech Ministry of Education, Youth and Sports. The work was
carried out partially within the framework of the European Associated Laboratory
``Astrophysics Poland--France'' and the COST Action MP0905  
``Black Holes in a Violent Universe''. VK and MD acknowledge the Czech Science
Foundation grants 205/07/0052 and 202/09/0772; TP acknowledges a postdoctoral 
grant No.\ 205/09/P468. 
\end{acknowledgements}

\end{document}